\documentclass[prl,showpacs,amsmath,amssymb,floatfix,twocolumn]{revtex4}

\usepackage{graphicx}
\usepackage{dcolumn}
\usepackage{bm}
\usepackage{psfrag}
\usepackage{subfigure}

\newcommand{\be}{\begin{eqnarray}}
\newcommand{\ee}{\end{eqnarray}}
\newcommand{\bn}{\begin{eqnarray*}}
\newcommand{\en}{\end{eqnarray*}}

\newcommand{\nn}{\nonumber \\}
\newcommand{\nl}{\\}

\renewcommand{\vec}[1]{\mbox{\boldmath$#1$}}
\renewcommand{\d}{\mbox{\rm d}}
\newcommand{\gslash}[1]{\mbox{\slash{\hspace{-2mm}}$#1$}}
\newcommand{\kbar}{\mathchar'26\mkern-9mu k}

\newcommand{\al}{\ensuremath{\alpha}}
\newcommand{\bt}{\ensuremath{\beta}}
\newcommand{\sg}{\ensuremath{\sigma}}
\newcommand{\gm}{\ensuremath{\gamma}}
\newcommand{\dl}{\ensuremath{\delta}}
\newcommand{\lm}{\ensuremath{\lambda}}

\newcommand{\gfive}{\ensuremath{\gm^5}}

\newcommand{\Gm}{\ensuremath{\Gamma}}

\newcommand{\ze}{\ensuremath{\hat{0}}}

\newcommand{\pvec}{\ensuremath{\vec{p}}}
\newcommand{\Pvec}{\ensuremath{\vec{P}}}
\newcommand{\nvec}{\ensuremath{\vec{n}}}

\newcommand{\Rvec}{\ensuremath{\vec{R}}}

\newcommand{\sgvec}{\ensuremath{\vec{\sg}}}

\newcommand{\Dvec}{\ensuremath{\vec{D}}}
\newcommand{\Jvec}{\ensuremath{\vec{J}}}

\newcommand{\Xvec}{\ensuremath{\vec{X}}}
\newcommand{\Wvec}{\ensuremath{\vec{W}}}
\newcommand{\Gmvec}{\ensuremath{\vec{\Gm}}}

\newcommand{\nabvec}{\ensuremath{\vec{\nabla}}}

\newcommand{\hb}{\ensuremath{\hbar}}

\newcommand{\lt}{\ensuremath{\left}}
\newcommand{\rt}{\ensuremath{\right}}

\renewcommand{\d}{\mbox{\rm d}}

\begin{document}







\title{Signatures of Noncommutative Geometry in Muon Decay for Nonsymmetric Gravity}%


\author{Dinesh Singh}
\altaffiliation[Electronic address:  ]{dinesh.singh@uregina.ca}
\author{Nader Mobed}
\altaffiliation[Electronic address:  ]{nader.mobed@uregina.ca}
\author{Pierre-Philippe Ouimet}
\altaffiliation[Electronic address:  ]{pierre-philippe.ouimet@uregina.ca}
\address{Department of Physics, University of Regina, Regina, Saskatchewan, S4S 0A2, Canada}

\date{\today}

\begin{abstract}
It is shown how to identify potential signatures of noncommutative geometry within the decay spectrum
of a muon in orbit near the event horizon of a microscopic Schwarzschild black hole.
This possibility follows from a re-interpretation of J.W.~Moffat's nonsymmetric theory of gravity, first
published in Phys. Rev. D {\bf 19}, 3554 (1979), where the antisymmetric part of the metric tensor manifests
the hypothesized noncommutative geometric structure throughout the manifold.
It is further shown that for a given sign convention, the predicted signatures counteract the effects of curvature-induced muon stabilization
predicted by D.~Singh and N.~Mobed in Phys. Rev. D {\bf 79}, 024026 (2009).
While it is unclear whether evidence for noncommutative geometry may be found at the Large Hadron Collider
(LHC) anytime soon, this approach at least provides a useful direction for future quantum gravity research
based on the ideas presented here.
\end{abstract}


\maketitle

{\sl Introduction.--}Efforts to find a unified 
theory of particle physics beyond the Standard Model (SM) have led researchers to consider 
various approaches towards reaching this goal.
One possibility that has gained widespread attention over the past two decades is string theory \cite{Polchinski},
primarily because of the claim that it includes a quantum theory of gravity within its framework.
However, there are numerous competing formulations, such as loop quantum gravity \cite{Ashtekar} and others,
with radically different foundations and implications as they relate to Beyond-Standard-Model (BSM) physics.
This is due to the fact that the SM offers no insights about how gravity behaves at quantum length scales.

A recent approach in quantum gravity of serious consideration is noncommutative geometry \cite{Connes},
which assumes that pairs of space-time co-ordinates violate commutation at some scale $\kbar$
with dimensions of area, 
and instead behave like noncommutating operators.
This property is described by 
\be
\lt[\Xvec^\mu , \Xvec^\nu\rt] & = & i \kbar \, \Jvec^{\mu \nu}(\Xvec^\al) \, ,
\label{noncomm}
\ee
where $\Xvec^\mu$ is a Hermitian co-ordinate operator and $\Jvec^{\mu \nu}$ is a dimensionless antisymmetric Hermitian 
operator, resembling orbital angular momentum generators with an SU(2) Lie algebra structure.
Historically, this proposal was introduced several decades ago \cite{Snyder} to truncate ultraviolet divergences
in the early developments of QED, but lost favour after successful techniques of regularization and renormalization
were introduced.
Nonetheless, noncommutative geometry has made a resurgence of late, in part because of its capacity
to offset curvature singularities inside black holes \cite{Nicolini} by truncating the curvature field strength
in much the same way as was done to overcome perceived deficiencies in quantum field theory.

In contrast to taking a direct approach for quantum gravity research, a recent investigation \cite{Singh-Mobed1}
focussed on the more intermediate 
problem of understanding the applicability of the
Poincar\'{e} group, the space-time symmetry group for classifying elementary particles according to their
mass and spin angular momentum, for situations involving the non-inertial motion of spin-1/2 particles
in space with respect to a global laboratory frame in a flat space-time background.
This is a relevant issue given that the Poincar\'{e} group is well-defined only for strictly inertial motion \cite{Singh-Mobed1},
leading to conceptual challenges when applied to an elementary particle's accelerated motion in flat space-time or
geodesic motion in curved space-time.
By describing the Poincar\'{e} group generators for the canonical momentum $\Pvec$ in terms of
curvilinear spatial co-ordinates to accommodate the symmetries of the particle's classical spatial trajectory,
it is shown that the corresponding Pauli-Lubanski spin four-vector
$\Wvec$ yields a Casimir scalar $\Wvec \cdot \Wvec$ that is no longer the Lorentz invariant
${1 \over 2}\lt({1 \over 2} + 1\rt) m_0^2$, where $m_0^2 = \Pvec \cdot \Pvec$.
Rather, $\Wvec \cdot \Wvec$ is now {\em frame dependent} due to an additive term coupling the Pauli spin operator $\sgvec$ with
a Hermitian three-vector $\Rvec^i = \lt(i \over 2\hbar \rt) \epsilon^{ijk} [\Pvec_j, \Pvec_k]$, known as
the {\em non-inertial dipole operator} \cite{Singh-Mobed1,Singh1} because it generates an interaction term analogous
to a dipole interaction with a magnetic field.
For $r$ the particle trajectory's local radius of curvature with respect to the laboratory frame,
the dimensionality of $\Rvec$ is $|\Pvec|/r$, with the property that $\Rvec \rightarrow \vec{0}$
for fixed momentum as $r \rightarrow \infty$, while for Cartesian co-ordinates to represent strictly rectilinear motion,
$\Rvec = \vec{0}$ identically.

More recently, this computation was repeated in the presence of a curved space-time background,
described in terms of Fermi normal co-ordinates.
Upon applying the formalism to the specific example of muon decay while orbiting a microscopic Kerr black hole \cite{Singh-Mobed2},
it was shown that the muon decay spectrum becomes significantly distorted due to both non-inertial effects from $\Rvec$
and curvature-induced contributions, ultimately leading to a predicted stabilization of the muon when approaching an orbital
radius close to the particle's Compton wavelength.
This by itself is a highly significant result, since it suggests that the intersection between quantum mechanics
and gravitation becomes relevant at a length scale over {\em 20 orders of magnitude} larger than the Planck scale
for this application, and suggests non-trivial behaviour of elementary particles in curved space-time
apparently unaccounted for by current explorations of quantum gravity research.
It also corroborates an earlier claim \cite{Rosquist} that predicts curvature-induced changes to the electric field
at the Compton scale generated by a charged microscopic Kerr black hole with the same mass and spin as an electron.
In addition, this paper noted an interesting observation that if the metric tensor
in Fermi normal co-ordinates ${}^F{}g_{\mu \nu}(X)$ were allowed to become {\em nonsymmetric} while simultaneously
allowing for $X^\mu \rightarrow \Xvec^\mu$ in accordance with (\ref{noncomm}), then it is theoretically possible to
identify non-trivial signatures of noncommutative geometry in the muon decay spectrum.
However, a preliminary computation with this extension yielded no contribution whatsoever due to noncommutative geometry,
a highly surprising result with no obvious explanation forthcoming.

Nonetheless, since this first investigation into noncommutative geometry, it was soon realized that a more complete computation
to yield a signature would require the full details of a suitably-defined nonsymmetric theory of gravity.
It so happens that a theory of nonsymmetric gravity developed years ago by Moffat \cite{Moffat1,Moffat2},
originally derived as a purely classical theory with a different motivation in mind,
may satisfy this requirement if the antisymmetric part of the metric tensor is identified exclusively with the noncommutative
geometric extension of space-time, as originally proposed \cite{Singh-Mobed2}.
The purpose of this paper is to present the outcome of this computation in a microscopic Schwarzschild space-time background
using Moffat's nonsymmetric gravity theory to incorporate additional curvature terms coupled to $\kbar$
that were previously not considered.
As well, this paper compares two possible formulations of noncommutative geometry, as determined by the choice for
$\Jvec^{\mu \nu}(\Xvec^\al)$ in (\ref{noncomm}).
While $\kbar$ may have theoretical physical constraints on its magnitude, in this paper
$\kbar$ is treated as a free parameter to be determined by observation.
Geometric units of $G = c = 1$ are assumed throughout, where the curvature tensors presented here satisfy the conventions of
MTW \cite{MTW} but with $-2$ metric signature.

{\sl Formalism.--}While details of the formalism employed here are given elsewhere \cite{Singh-Mobed2},
it is important to offer some relevant highlights to justify its later extension to include noncommutative geometry.
Given Fermi normal co-ordinates $X^\mu = \lt(T, X^i\rt)$ defined in a local neighbourhood 
about a spin-1/2 particle's worldline, where $T$ is the proper time and
$X^i$ is the local Cartesian spatial co-ordinate in the Fermi frame, the covariant Dirac equation with mass $m$ and
$\partial_\mu = \partial/\partial X^\mu$ is
\be
\lt[i \gm^\mu(X) \lt(\partial_\mu + i \, \Gamma_\mu(X) \rt) - m/\hb\rt]\psi(X) & = & 0 \, .
\label{Dirac-eq}
\ee
The set of gamma matrices $\lt\{ \gm^\mu(X) \rt\}$ satisfy
$\lt\{ \gm^\mu(X), \gm^\nu(X) \rt\} = 2 \, g_F^{\mu \nu}(X)$ and $\Gamma_\mu(X)$ is the spin connection.
An orthonormal vierbein set $\lt\{ \bar{e}^\mu{}_{\hat{\alpha}}(X) \rt\}$ and its inverse
$\lt\{ \bar{e}^{\hat{\alpha}}{}_\mu (X)\rt\} \,$ can be obtained to define a local Lorentz frame, 
denoted by co-ordinates with hatted indices.
The metric tensor is then described by
${}^F{}g_{\mu \nu}(X) = \eta_{\hat{\alpha}\hat{\beta}} \, \bar{e}^{\hat{\alpha}}{}_\mu (X) \, \bar{e}^{\hat{\beta}}{}_\nu (X)$,
such that
\begin{subequations}
\label{F-g}
\be
{}^F{}g_{00}(X) & = & 1 - {}^F{}R_{l00m}(T) \, X^l \, X^m + \cdots \, ,
\label{F-g00}
\nl
{}^F{}g_{0j}(X) & = & -{2 \over 3} \, {}^F{}R_{l0jm}(T) \, X^l \, X^m + \cdots \, ,
\label{F-g0j}
\nl
{}^F{}g_{ij}(X) & = & \eta_{ij} - {1 \over 3} \, {}^F{}R_{lijm}(T) \, X^l \, X^m + \cdots \, ,
\label{F-gij}
\ee
\end{subequations}
where $\eta_{\mu \nu}$ is the Minkowski metric and ${}^F{}R_{\mu \al \bt \nu}(T)$ is the projection of the Riemann tensor
for general relativity (GR) onto the Fermi frame. 

A straightforward conversion of (\ref{Dirac-eq}) to mutually orthogonal spatial curvilinear co-ordinates is then introduced,
defined by $U^\mu = \lt(T, u^i\rt)$, where $X^i = X^i(u)$.
Subsequent projection of (\ref{Dirac-eq}) onto the local Lorentz frame leads to
\be
\lt[\gm^{\hat{\mu}} \lt(\Pvec_{\hat{\mu}} - \hb \, \Gmvec_{\hat{\mu}}(U) \rt) - m\rt]\psi(U) & = & 0 \, ,
\label{Dirac-eq-curvilinear-1}
\ee
where
\be
i \, \Gmvec_{\hat{\mu}} & = & \bar{\Gmvec}^{(\rm S)}_{\hat{\mu}}
+ \gm^{\hat{l}} \, \gm^{\hat{m}} \, \bar{\Gmvec}^{(\rm T)}_{\ze[\hat{l}\hat{m}]} \, \dl^{\ze}{}_{\hat{\mu}} \, ,
\label{Spin-Connection}
\ee
and
\begin{subequations}
\be
\bar{\Gmvec}^{(\rm S)}_{\ze} & = & {1 \over 12} {}^F{}R^m{}_{jmk,0}(T) \, X^j \, X^k \, ,
\label{Spin-Connection-0-S}
\nl
\bar{\Gmvec}^{(\rm S)}_{\hat{\jmath}} & = & - \lt[{1 \over 2} {}^F{}R_{j00m}(T) + {1 \over 3} {}^F{}R_{jl0m}(T) \, X^l \rt] X^m \, ,
\qquad \hspace{1mm}
\label{Spin-Connection-j-S}
\nl
\bar{\Gmvec}^{(\rm T)}_{\ze[\hat{l}\hat{m}]} & = & {1 \over 12} \, {}^F{}R_{k[lm]j,0}(T) \, X^j \, X^k
\nn
& &{} + {1 \over 3} \, \lt[{}^F{}R_{lm0k}(T) + {}^F{}R_{k[lm]0}(T)\rt] X^k \, .
\label{Spin-Connection-0-T}
\ee
\end{subequations}
After defining $\Dvec_{\hat{\mu}} = \Pvec_{\hat{\mu}} - \hb \, \Gmvec_{\hat{\mu}}$ and using the identity 
\be
\gm^{\hat{\mu}} \, \gm^{\hat{\nu}} \, \gm^{\hat{\rho}} & = & \eta^{\hat{\nu} \hat{\rho}} \, \gm^{\hat{\mu}}
- 2 \, \gm^{[\hat{\nu}} \eta^{\hat{\rho}]\hat{\mu}} - i \, \gm^5 \, \gm^{\hat{\sg}} \, \varepsilon^{\hat{\mu} \hat{\nu} \hat{\rho}}{}_{\hat{\sg}} \, ,
\label{gm-identity}
\ee
where $\varepsilon^{\hat{\mu} \hat{\nu} \hat{\rho} \hat{\sg}}$ is the Levi-Civita symbol with
$\varepsilon^{\hat{0} \hat{1} \hat{2} \hat{3}} = 1$, it follows that
\be
\Dvec_{\hat{\mu}} & = & \Pvec_{\hat{\mu}} -
\hb \lt(\gm^5 \, \bar{\Gmvec}^{(\rm C)}_{\hat{\mu}} - i \, \bar{\Gmvec}^{(\rm S)}_{\hat{\mu}}\rt) \, ,
\label{D}
\nl
\bar{\Gmvec}^{(\rm C)}_{\hat{\mu}} & = & \varepsilon^{\hat{0} \hat{l} \hat{m}}{}_{\hat{\mu}} \, \bar{\Gmvec}^{(\rm T)}_{\ze[\hat{l}\hat{m}]} \, ,
\label{Spin-Connection-C}
\ee
where ``S'' gives the symmetric part of the spin connection under chiral symmetry and ``C'' implies chiral dependence
by its coupling with $\gm^5$.

Following standard definitions, the Pauli-Lubanski vector in a local Lorentz frame is
\be
\Wvec^{\hat{\mu}} & = & -{1 \over 4} \, \varepsilon^{\hat{\mu}}{}_{\hat{\al}\hat{\bt}\hat{\gm}} \,
\sg^{\hat{\al}\hat{\bt}} \, \Dvec^{\hat{\gm}} \, .
\label{Wvec-grav-em}
\ee
Using both (\ref{gm-identity}) and the identity
$\varepsilon_{\hat{\mu} \hat{\nu} \hat{\rho} \hat{\sg}} \, \sg^{\hat{\rho} \hat{\sg}}
= -2 \, i \, \gm^5 \, \sg_{\hat{\mu} \hat{\nu}}$, 
its squared magnitude is
\be
\Wvec^{\hat{\al}} \, \Wvec_{\hat{\al}} & = &
-{3 \over 4} \, \Dvec^{\hat{\al}} \, \Dvec_{\hat{\al}} 
+ {i \over 4} \, \sg^{\hat{\al}\hat{\bt}} \lt[\Dvec_{\hat{\al}} , \Dvec_{\hat{\bt}}\rt] \,  \hspace{1mm}
\label{W^2-1}
\ee
and leads to 
\be
\Wvec^{\hat{\al}} \, \Wvec_{\hat{\al}} & = & -{3 \over 4} \lt[m_0^2 + {\hbar^2 \over 6} \lt({}^F{}R^{\hat{\al}}{}_{\hat{\al}}\rt) \rt]
+ {\hbar \over 2} \lt(\sgvec \cdot \Rvec\rt)
\nn
&  &{} -{\hbar \over 4} \, \sg^{\hat{\al} \hat{\bt}} \, Q_{\hat{\al}\hat{\bt}}
+ {3 \over 2} \, \hbar \lt(\gm^5 \, \bar{\Gmvec}^{(\rm C)}_{\hat{\al}} - i \, \bar{\Gmvec}^{(\rm S)}_{\hat{\al}}\rt) \Pvec^{\hat{\al}}
\nn
&  &{} + {3 \over 4} \, \hbar^2 \, \nabvec^{\hat{\al}}
\lt(\bar{\Gmvec}^{(\rm S)}_{\hat{\al}} + i \, \gm^5 \, \bar{\Gmvec}^{(\rm C)}_{\hat{\al}}\rt) \, ,
\label{W^2-2}
\ee
where $m_0^2 = \Pvec^{\hat{\al}} \, \Pvec_{\hat{\al}}$, the Casimir invariant for momentum,
$\nabvec_{\hat{\mu}}$ is the gradient operator in curvilinear co-ordinates,
and $Q_{\hat{\al}\hat{\bt}}$ is purely curvature-dependent.
When ${}^F{}R_{\mu \nu \al \bt}(T) \rightarrow 0$, (\ref{W^2-2}) reduces to its flat space-time expression \cite{Singh-Mobed1},
where ${\hbar \over 2} \lt(\sgvec \cdot \Rvec\rt)$ is the non-inertial dipole interaction term
to justify the given name for $\Rvec$.

{\sl Extension for Noncommutative Geometry.--}To ensure that (\ref{F-g}) is truly symmetric requires
assuming that the Riemann tensor is also symmetric in the middle two indices.
Since this is generally not true, the symmetric properties of ${}^F{}g_{\mu \nu}(X)$ follow directly from
assuming the condition that $\d s^2 = {}^F{}g_{\mu \nu}(X) \, \d X^\mu \, \d X^\nu$,
where $X^\mu$ and $\d X^\mu$ are ordinary c-numbers.
This is an important observation given that
the {\em antisymmetric} combination ${}^F{}R_{\mu [\al \bt] \nu}(T) =
{1 \over 2} \lt[{}^F{}R_{\mu \al \bt \nu}(T) - {}^F{}R_{\mu \bt \al \nu}(T)\rt]$ leads to the appearance
of terms like
\be
{}^F{}R_{\mu [\al \bt] \nu}(T) \, X^\mu \, X^\nu & = & -{1 \over 4} \, {}^F{}R_{\al \bt \mu \nu}(T) \lt[X^\mu , X^\nu\rt] \, . \qquad
\label{Riemann-antisymmetric}
\ee

Assuming that $X^\mu$ are ordinary c-numbers, it follows that (\ref{Riemann-antisymmetric}) automatically vanishes.
However, if $X^\mu \rightarrow \Xvec^\mu$, then (\ref{Riemann-antisymmetric}) is nonzero from (\ref{noncomm}),
which results in $\d s^2 \rightarrow
{1 \over 2} \, {}^F{}g_{(\mu \nu)}(\Xvec) \lt( \d \Xvec^\mu \otimes \d \Xvec^\nu + \d \Xvec^\nu \otimes \d \Xvec^\mu \rt)
+ {1 \over 2} \, {}^F{}g_{[\mu \nu]}(\Xvec) \, \d \Xvec^\mu \wedge \d \Xvec^\nu$, where $ {}^F{}g_{[\mu \nu]}(\Xvec)$ is
a Hermitian operator defined at the $\kbar$-scale, and
\be
{}^F{}g_{[\mu j]}(\Xvec) & = & {i \kbar \over 6} \lt[\dl^0{}_\mu {}^F{}R_{0jkl}(T) + {1 \over 2} \, \dl^i{}_\mu {}^F{}R_{ijkl}(T)\rt]
\nn
&  &{} \times \Jvec^{kl}(\Xvec) \, .
\label{F-g-antisymm}
\ee
This leads to an immediate extension of the spin connection, in the form
\begin{subequations}
\label{spin-connection-noncomm}
\be
\bar{\Gmvec}^{(\rm S)}_{\hat{\mu}} & \rightarrow &
\bar{\Gmvec}^{(\rm S)}_{\hat{\mu}} + {i \kbar \over 12} \, {}^F{}R_{lm0\mu,0}(T) \, \Jvec^{lm} \, , \qquad
\label{Spin-Connection-mu-S-noncomm}
\nl
\bar{\Gmvec}^{(\rm C)}_{\hat{\mu}} & \rightarrow &
\bar{\Gmvec}^{(\rm C)}_{\hat{\mu}} + {i \kbar \over 48} \, \varepsilon^{\hat{0} \hat{\jmath} \hat{k}}{}_{\hat{\mu}}
\, {}^F{}R_{lmjk,0}(T) \, \Jvec^{lm} \, ,
\label{Spin-Connection-mu-C-noncomm}
\ee
\end{subequations}
which subsequently introduces non-trivial terms to (\ref{Wvec-grav-em}).

{\sl Moffat's Nonsymmetric Gravity Theory.--}It is reasonable to surmise that metric corrections
due to noncommutative geometry, as proposed by (\ref{F-g-antisymm}), are very small.
Nevertheless, by allowing for ${}^F{}g_{[\mu \nu]}(\Xvec) \neq 0$,
it is clear that a generalization away from GR is still required to accommodate a nonsymmetric description
of gravity.
The theory put forward by Moffat \cite{Moffat1,Moffat2} proposes that both the metric tensor $g_{\mu \nu}$ and
the connection $\Gm^\lm{}_{\mu \nu}$ are nonsymmetric in their lower indices.
Furthermore, the modified curvature tensor is described using the nonsymmetric connection
\be
W^\lm{}_{\mu \nu} & = & \Gm^\lm{}_{\mu \nu} - {2 \over 3} \, \dl^\lm{}_\mu \, W_\nu \, ,
\label{W-connection}
\ee
where $W_\mu = W^\lm{}_{[\mu \lm]}$ is a vector field expressed in terms of $g_{[\mu \nu]}$.
The modified tensor components in the Fermi frame are then introduced into the
orthonormal vierbeins $\lt\{ \bar{e}^\mu{}_{\hat{\alpha}}(X) \rt\}$ and inverse vierbeins
$\lt\{ \bar{e}^{\hat{\alpha}}{}_\mu (X)\rt\}$ \cite{Singh-Mobed2}.
Moffat's theory is particularly useful in the linear approximation about the Minkowski metric \cite{Moffat2},
since the symmetric and antisymmetric field equations decouple to first order. 
Given that ${}^F{}g_{\mu \nu}(X)$ according to (\ref{F-g}) is expressible precisely as a linearized expansion about $\eta_{\mu \nu}$,
this particular form of Moffat's nonsymmetric gravity theory is especially useful in this context.

\begin{figure*}
\psfrag{x}[tc][][1.8][0]{\Large $x$}
\psfrag{dG}[bc][][1.8][0]{\Large ${1 \over \Gm_0} \, \hat{\Gm}(x)$}
\vspace{1mm}
\begin{minipage}[t]{0.3 \textwidth}
\centering
\subfigure[\hspace{0.2cm} $M = 3 \times 10^{-12}$ cm ($r_0 \sim 6 M$)]{
\label{fig:Michel-background}
\rotatebox{0}{\includegraphics[width = 5.0cm, height = 3.1cm, scale = 1]{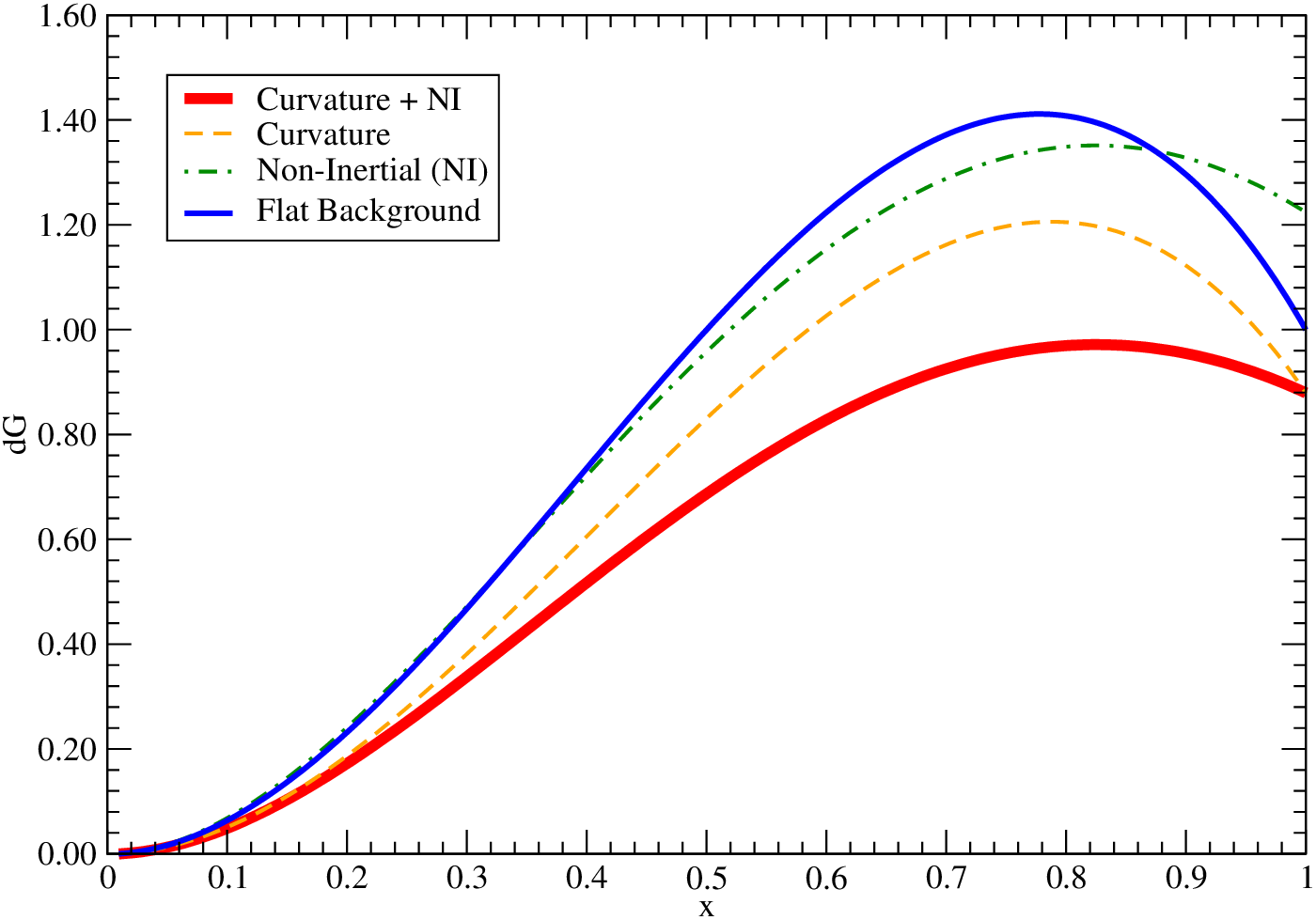}}}
\end{minipage}%
\hspace{0.5cm}
\begin{minipage}[t]{0.3 \textwidth}
\centering
\subfigure[\hspace{0.2cm} $\sqrt{\kbar} = 5 \times 10^{-9}$ cm (Case 1)]{
\label{fig:Michel-Case-1}
\rotatebox{0}{\includegraphics[width = 5.0cm, height = 3.1cm, scale = 1]{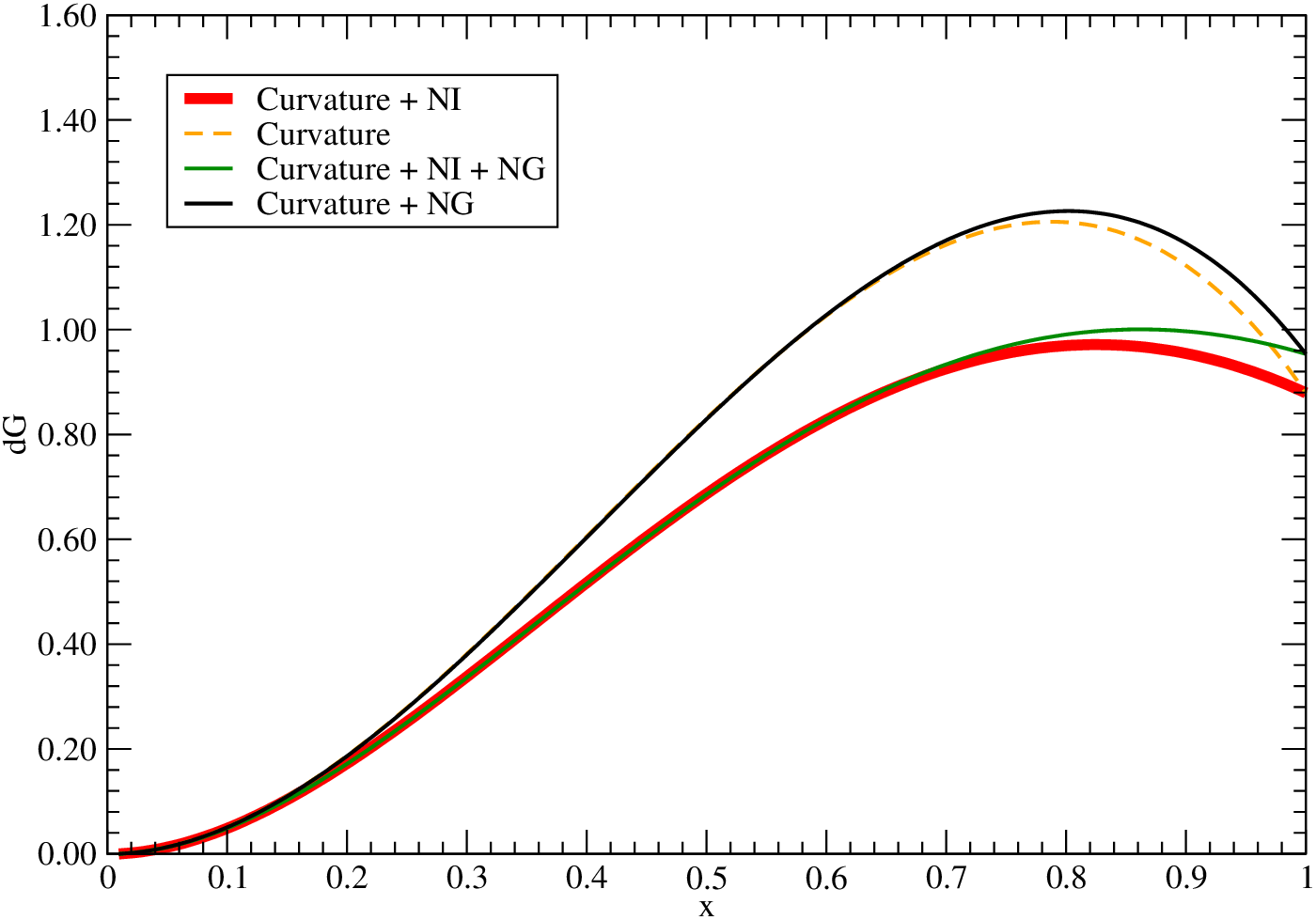}}}
\end{minipage}%
\hspace{0.5cm}
\begin{minipage}[t]{0.3 \textwidth}
\centering
\subfigure[\hspace{0.2cm} $\sqrt{\kbar} = 2 \times 10^{-8}$ cm (Case 2)]{
\label{fig:Michel-Case-2}
\rotatebox{0}{\includegraphics[width = 5.0cm, height = 3.1cm, scale = 1]{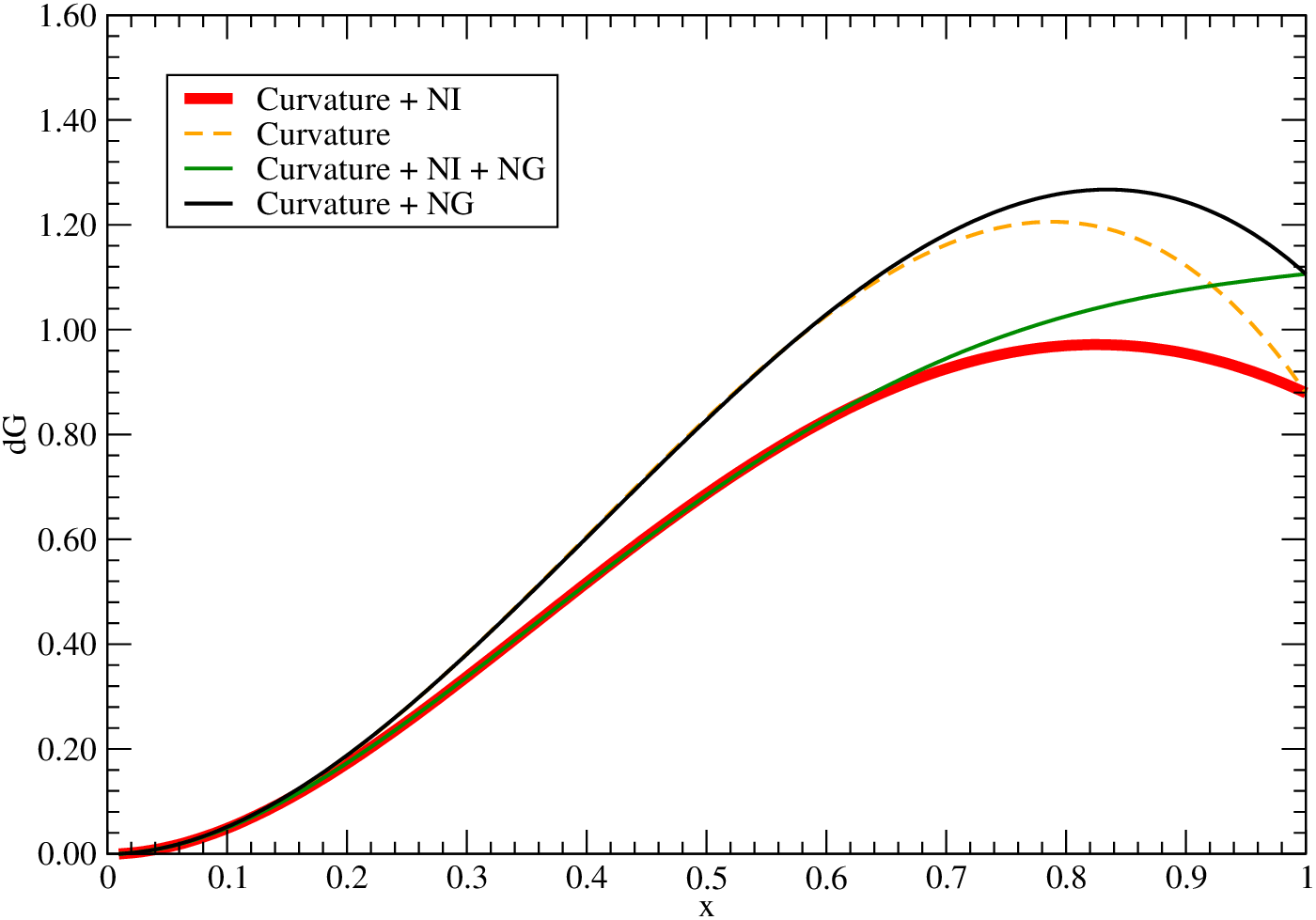}}}
\end{minipage}
\caption{\label{fig:Michel-spectra} Michel spectrum for muon decay while in circular orbit around a microscopic Schwarzschild black hole
of mass $M = 3 \times 10^{-12}$ cm and orbital radius $r_0 \sim 6 M$.
Fig.~\ref{fig:Michel-background} shows the various profiles available,
including both curvature and $\Rvec$-dependent non-inertial contributions (NI) \cite{Singh-Mobed2}, minus the predictions due to
noncommutative geometry (NG).
Fig.~\ref{fig:Michel-Case-1} shows the contributions due to noncommutative geometry for Case 1, given by (\ref{J-structure-const-1}),
while Fig.~\ref{fig:Michel-Case-2} shows the corresponding description for Case 2, given by (\ref{J-structure-const-2}).
\vspace{1mm}}
\end{figure*}
{\sl Noncommutative Geometry in Muon Decay.--}For this paper, the decay process occurs
for a muon in circular orbit around a microscopic Schwarzschild black hole \cite{Singh-Mobed2}
of mass $M = 3 \times 10^{-12}$ cm with (classical) orbital radius $r_0 \sim 6 M$, where $X^j$ represents spatial quantum fluctuations
about $r_0$.
The matrix element \cite{Singh-Mobed1,Singh-Mobed2} for the muon decay reaction $\mu^- \rightarrow e^- + \bar{\nu}_e + \nu_\mu$ is
\be
|{\cal M}|^2 & = &
{G_{\rm F}^2 \over 2} \, L^{(\mu)}_{\hat{\mu} \hat{\nu}} \, M^{\hat{\mu} \hat{\nu}}_{(e)},
\label{M^2}
\ee
where
%
\begin{subequations}
\label{L-M}
\be
\hspace{-0.8cm}
L^{\hat{\mu} \hat{\nu}}_{(\mu)} & = & {\rm Tr} \lt[\gslash{\pvec}_{\nu_\mu} \, \gm^{\hat{\mu}} \lt(\gslash{\Dvec}_{\mu}
+ m_{\mu} \, \gfive \, \gslash{\nvec}_{\mu}\rt)\gm^{\hat{\nu}} \lt(1 - \gfive\rt)\rt] \, ,
\label{L_uv}
\nl
\hspace{-0.8cm}
M^{\hat{\mu} \hat{\nu}}_{(e)} & = & {\rm Tr} \lt[\lt(\gslash{\Dvec}_{e} + m_{e} \, \gfive \, \gslash{\nvec}_{e}\rt)
\gm^{\hat{\mu}} \, \gslash{\pvec}_{\nu_e} \, \gm^{\hat{\nu}} \lt(1 - \gfive\rt)\rt] \, ,
\label{M_uv}
\ee
\end{subequations}
and $\nvec^{\hat{\mu}}$ is the polarization vector for the charged lepton.
From the orthogonality condition $\nvec^{\hat{\mu}} \, \Dvec_{\hat{\mu}} = 0$, it can be shown \cite{Singh-Mobed2} that
$\gslash{\nvec} = \lt[\gslash{\nvec^{\rm (S, Re)}} + i \, \gslash{\nvec^{\rm (S, Im)}}\rt]
+ \gm^5 \lt[\gslash{\nvec^{\rm (C, Re)}} + i \, \gslash{\nvec^{\rm (C, Im)}}\rt]$, eventually leading
to
\be
|{\cal M}|^2 & = & 32 \, G_{\rm F}^2 \lt(\Pvec_{\nu_e}^{\hat{\al}} \, \bar{\Dvec}^{\mu}_{\hat{\al}} \rt)
\lt(\Pvec_{\nu_\mu}^{\hat{\bt}} \, \bar{\Dvec}^{e}_{\hat{\bt}} \rt) \, ,
\label{M^2=}
\ee
where $\bar{\Dvec}_{\hat{\al}} = \Pvec_{\hat{\al}} - \hbar \lt(\bar{\Gmvec}^{\rm (C)}_{\hat{\al}} - i \, \bar{\Gmvec}^{\rm (S)}_{\hat{\al}} \rt)
+ m_0 \lt(\nvec_{\hat{\al}}^{\rm (C)} - \nvec_{\hat{\al}}^{\rm (S)}\rt)$.
Both the gravitational and $\Rvec$-dependent contributions to the muon decay rate are additive corrections to
$\Gamma_0 \approx G_{\rm F}^2 \, m_\mu^5/(192 \, \pi^3) \approx 2.965 \times 10^{-16}$~MeV \cite{Singh-Mobed2}.
As well, $|{\cal M}|^2$ formally contains new terms due to noncommutative geometry when accounting for
(\ref{spin-connection-noncomm}) and the additional curvature terms due to Moffat's theory.

For this paper, $\Jvec^{\mu \nu}$ is defined according to two possible choices \cite{Madore}:
\begin{subequations}
\label{J-structure-const}
\be
\Jvec^{\mu \nu}(\Xvec) & = & i \, C_{(0)}^{\mu \nu} \, , \qquad \qquad {\rm (Case \ 1)}
\label{J-structure-const-1}
\nl
& = & {i \over r_0} \, C_{(1)}^{\mu \nu}{}_\gm \, \Xvec^\gm \, , \quad {\rm (Case \ 2)}
\label{J-structure-const-2}
\ee
\end{subequations}
where $C_{(0)}^{\mu \nu}$ and $C_{(1)}^{\mu \nu}{}_\gm$ are dimensionless real-valued structure constants
chosen with $C_{(0)}^{ij} = C_{(1)}^{ij}{}_k = +1$ for $ij = \{12, 23, 31\}$, and all others equal to zero.
When applied to Moffat's nonsymmetric theory \cite{Moffat2}, it is shown that the components for $W_\mu$
(as c-numbers) in the Fermi frame for Case 1 are
%
\label{W-Case-1}
\be
{}^F{}W_\mu(X) & = & {\kbar \over 3} \, {}^F{}R_{kl0\mu,0}(T) \, C_{(0)}^{kl} \, ,
\label{Wj-Case-1}
\ee
%
while the corresponding components for Case 2 are
\begin{subequations}
\label{W-Case-2}
\be
{}^F{}W_0(X) & = & {\kbar \over 3 \, r_0} \, {}^F{}R_{klm0}(T) \, C_{(1)}^{klm} \, ,
\label{W0-Case-2}
\nl
{}^F{}W_j(X) & = & {\kbar \over 3 \, r_0} \lt[{}^F{}R_{kl0j,0}(T) \, C_{(1)}^{kl}{}_m \, X^m \rt.
\nn
&  &{} + \lt. {1 \over 2} \, {}^F{}R_{klmj}(T) \, C_{(1)}^{klm} \rt] \, .
\label{Wj-Case-2}
\ee
\end{subequations}

{\sl Analysis and Conclusion.--}All computations in this paper are performed in the muon's local rest frame.
Figure~\ref{fig:Michel-spectra} shows the Michel spectrum $\hat{\Gm}(x) = \d \Gm/\d x$ in
units of $\Gm_0$ as a function of the outgoing electron energy fraction $x$.
It is clear from Fig.~\ref{fig:Michel-background} in the absence of noncommutative geometry
that space-time curvature reduces the decay rate \cite{Singh-Mobed2}.
Figs.~\ref{fig:Michel-Case-1} and \ref{fig:Michel-Case-2} reveal the presence of $\kbar$
due to Cases 1 and 2, respectively, where the change of profile occurs for $x \gtrsim 0.65$.
It is interesting to note that this contribution serves to counteract
the emergent curvature-induced stabilization at the muon's Compton wavelength scale for the adopted sign
convention of the structure constants.
As well, it is surprising to note that a signature appears only for terms {\em quadratic} in curvature
but linear in $\kbar$, as all such linear curvature terms identically {\em vanish} when time-averaged over an orbital cycle.
The reasons for its occurrence are not known.

Identification of a discernable signal requires $\sqrt{\kbar} \sim 10^{-9}$ cm for Case 1, about
an order of magnitude smaller than for Case 2.
While this length scale for $\sqrt{\kbar}$ is within the realm of possibility and is strictly speaking
an unknown quantity, its value is at least ten orders of magnitude larger than the theoretical upper bound of
$\sqrt{\kbar} \sim 10^{-18}$~cm \cite{Chaichian} predicted in the context of noncommutative QED in flat space-time.
Since this treatment of noncommutative geometry requires a curved space-time background to manifest its existence,
it is unclear how the empirically-motivated choices for $\sqrt{\kbar}$ in Figs.~\ref{fig:Michel-Case-1} and \ref{fig:Michel-Case-2}
relate with the theoretical motivations given elsewhere.
Irrespective of whether the Large Hadron Collider (LHC) can produce Schwarzschild black holes
without invoking large extra spatial dimensions, it is also unclear whether direct observational evidence for
noncommutative geometry can be gained by this method.
Nonetheless, this approach at least provides an interesting avenue for further exploration.



\end{document}